\documentclass[12pt]{article}
\usepackage[english]{babel}
\usepackage[dvips]{graphicx}

\title{\bf PHOTON-NEUTRINO INTERACTIONS}
\author{G.Karl and V.Novikov\footnote{On leave of absence from the Institute of Theoretical and Experimental Physics, Moscow, Russia}, \\ Department of Physics, University of Guelph,\\ Ontario, Canada N1G 2W1\\ and Guelph-Waterloo Physics Institute}
\date{}

\def\fun#1#2{\lower3.6pt\vbox{\baselineskip0pt\lineskip.9pt
\ialign{$\mathsurround=0pt#1\hfil##\hfil$\crcr#2\crcr\sim\crcr}}}

\begin{document}
\maketitle

\begin{abstract}
 We discuss the interaction of photons with neutrinos
including two lepton loops. The parity violation in the 
$\gamma \nu \to \gamma \nu$ channel due to two lepton loops is
substantially enhanced relative to the one lepton loop contribution.
However there is no corresponding enhancement in the parity conserving
amplitude in either the direct or the cross channel
$\gamma \gamma \to \nu\bar{\nu}$.

\end{abstract}

\vspace {5mm}

\section  {Introduction and Summary}

The photon-neutrino interaction is very feeble, involving only
electrically neutral external particles. The cross sections are exceedingly tiny. Therefore this interaction can only be of astrophysical interest. Chiu and Morrison \cite{1} proposed long ago that
photon conversion to neutrinos may play a role in supernova cooling.
Another possible application involves the propagation of light waves
through a handed neutrino sea \cite{2} which  results in optical activity
(birefringence). There may perhaps be effects in the cooling of the
early universe. For all these reasons there has been some small
theoretical interest in this interaction \cite{3}. We consider here the
interaction at energies which are small compared to the mass of the weak bosons. At these energies the weak interaction can be described by a Fermi type effective theory. To have $\nu\gamma$ interaction one needs virtual leptons which couple both weakly and electromagnetically, thus lepton  loops in Feynman graphs. Contrary to intuition a two loop graph
dominates the parity violation. In this section we summarize our results using only dimensional arguments. A more technical description is in the next sections.

 For supernova cooling the amplitude for the annihilation $\gamma \gamma \to \nu\bar{\nu}$      is
relevant, while for photon propagation in a neutrino sea the parity
violation in the cross(scattering) channel $\gamma \nu \to \gamma \nu$       is important. We discuss  first the annihilation process where the
two loop contribution turns out to be small. It is natural to expect
higher order graphs to be smaller
but as mentioned we find a violation of this rule here.  This is
reminiscent of a surprise in coherent photon scattering from atoms
discussed by G.E.Brown and Woodward and later Peierls 
\cite{4}.

The earliest estimate for the annihilation amplitude used the Fermi
theory of weak interactions \cite{1} , but  Gell-Mann \cite{5} noted that in $V-A$
theory  this amplitude vanishes for a point interaction of four
fermions. This can be understood physically without any detailed
computation. In Fermi Theory the annihilation amplitude is described by
a triangle graph, with an electron running in the triangular loop (see Fig.1).

\begin{figure}[h]
\begin{center}
\includegraphics[width=5cm]
{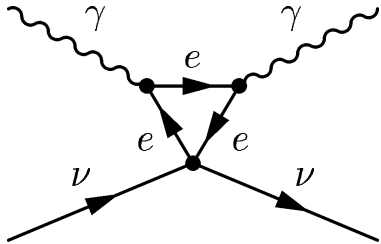}
\end{center}
\caption{$\gamma\nu$ scattering in the four-fermion Model.}
\label {Fig:1}
\end{figure}

 Two of the vertices of the triangle are electromagnetic (where the two photons couple) and the third vertex is weak, where the two neutrinos
emerge. The amplitude has magnitude $G_F\alpha (pk)$, where $p$ and $k$ are the
four-momenta of the neutrino and photon, and $pk$ is the only
non-vanishing relativistic invariant for the process (the Feynman amplitude is dimensionless which brings the factor $pk$). Because the two
neutrinos emerge from the same space-time point they can have no
relative orbital angular momentum (s-wave) .  With vector or axial
vector coupling ($V-A$), the spins of the fermion-anti-fermion pair must be
parallel. A theorem of Landau and Yang \cite{6} forbids two free photons in
a state with total angular momentum one. So the process is forbidden at
order $G_F$  in    $V-A$ theory. Gell-Mann also noted that this proof does
not hold if there is a weak boson mediating the weak interaction, so
that the two neutrinos are emitted at separate space-time points.
  
With a weak boson to mediate the interaction, the (parity conserving)
amplitude was estimated by Levine \cite{7}, and his estimate remains valid in
the standard model, as noted by a number of authors \cite{3, 8}.With the weak boson the loop changes from a triangle to a square, with the weak boson
providing one of the sides (see Fig.2).

\begin{figure}[h]
\begin{center}

\includegraphics{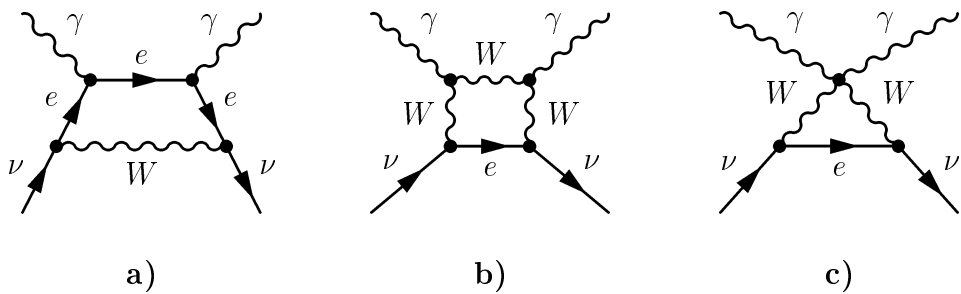}
\caption{ $\gamma\nu$ scattering in the Standard Model.}
\end{center}
\label{fig2}
 \end{figure}

 This gives an additional factor of
$(pk/M^2_W)$ to first order in the momenta $p$ and $k$. This suppression factor
vanishes when the boson mass goes to infinity, in agreement with
Gell-Mann's theorem. The precise cross
section for the  annihilation was computed by Dicus and Repko \cite{8} to order $(G_F \alpha)^2$ :

\begin {equation}
 d\sigma /dz = [(G_F\alpha)^2/32\pi^3][ \omega^6/M^4_W][1-z^4] \;\;,
\label{1}
\end{equation}
where $z$ is $cos\theta$  and $\omega$  the photon energy(in CoM system: $p k =\omega^2 (1-z)$)
and we have dropped  logarithms. This is in agreement with the one loop
Feynman amplitude described above, up to a constant factor:
\begin {equation}
 T^{(1)} = [G_F\alpha\omega^4/M^2_W][angular function][logarithms]
\;\;.
\label{2}
\end{equation}
The suppression factor ($\omega^2 /M^2_W$) may be interpreted as a $(kR)^2$ factor
associated with a d-wave for the two neutrinos. Therefore the leading
order amplitude may be thought of as anomalously small compared to
initial expectations \cite{8} . 

When we go to the next order in weak
interactions there is a set of diagrams in which the two neutrinos are
emitted at a separation of the order of the Compton wavelength of the
electron instead of the Compton wavelength of the weak boson. The
diagram remains a square with two adjacent  weak (order $G_F$) vertices, separated by an  electron-neutrino loop. The two final neutrinos are emitted at the two weak vertices, and the two photons couple at the two remaining vertices (see Fig.3).

\begin{figure}[h]
\begin{center}
\includegraphics[width=3cm]
{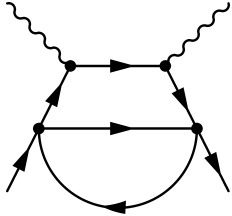}
\end{center}
\caption{The sample of two-loop contribution into $\gamma\nu$  scattering}
\label {Fig:3}
\end{figure}

 We can estimate the Feynman amplitude associated with the two-loop graph in lowest order using the same dimensional
arguments as in the case of the one-loop result, changing the $M_W$ with $m_e$ in the denominator:
\begin{equation}
T^{(2)} = [G^2_F \alpha (pk)^3/m^2_e] = [G^2_F\alpha \omega^  6/m^2_e][angular function ]
\;\;
\label{3}
\end{equation}
and therefore the ratio of the two-loop amplitude to the one loop is:
\begin{equation}
  T^{(2)}/T^{(1)}= G_F\omega^2(M^2_W/m^2_e)
\;\;.
\label{4}
\end{equation}

The bracketed ratio of the two squared masses is very large $\sim 10^{10}$ so
that
the cross-section is dominated by the leading term $T^{(1)}$ only at low energies
$(\omega< m_e)$. However at intermediate photon energies $\omega > m_e$ our estimate of
the two loop contribution is no longer valid since the ratio $(pk)/m^2_e$
is no longer small. To estimate what happens at these energies we take
the limit where this ratio is large and then we expect $m^2_e$ to be
replaced by $(pk)$ . Then the ratio of the two loop contribution to the one
loop contribution becomes of order $G_F M^2_W$   which is a small number (of
order $10^{-2}$ ) independent of energy.
Therefore there is no enhancement of the annihilation amplitude due to
two loop diagrams. This is in accord with expectations.
 
We come now to the parity violating amplitude in the cross channel.
This was first studied by Royer  \cite{2}, using the triangle graph, and his
result criticized by Stodolsky  \cite{2} on the basis of Gell-Mann's theorem.
More recently estimates have been proposed using the standard model  \cite{3}
and the one lepton loop graph, for both real and virtual photons. Here
we show that these one loop estimates are all negligible when compared
to the two lepton loop estimate-for real photons.
  
In the cross channel the \underline {parity violating} amplitude has to be odd under
inversion of the coordinates. The precise form is discussed in section 2, and given below. As a result, parity violation appears dimensionally as a two loop effect even
for a one loop amplitude. So it is natural to  consider two loop terms
together with one loop terms.

The estimate of the parity violating Feynman amplitude in the standard
model can be written in the form  \cite{3}:
\begin {equation}
T^{(1)}_{pv} =C\frac{G_F\alpha(pk)^2}{\sqrt{2}M^2_W}\epsilon_{\mu\nu\alpha\beta} \epsilon_{\mu}(k)
\epsilon_{\nu}^*(k) (\frac{p_{\alpha} k_{\beta}}{M^2_W})
\sim 194 G_F \alpha \omega^2 (\frac{\omega^4}{M^4_W}),
\;\;
\label{5}
\end {equation}
where $C$ is a constant involving logarithms  \cite{3} and in the second line we
evaluate the magnitude of the first expression in the forward direction
for photons of energy $\omega$   where $pk = 2\omega^2$. This formula should apply for $\omega < M_W$.

 With two lepton loops the corresponding parity violating Feynman amplitude has the form(for $\omega<m_e$)
\begin {equation}
 T^{(2)}_{pv} = \frac{3 G^2_F\alpha(pk)^2}{32\pi^3} \epsilon_{\mu\nu\alpha\beta} \epsilon_{\mu}(k)
\epsilon_{\nu}^*(k) [\frac{p_{\alpha} k_{\beta}}{m^2_e}]\sim
 0.025G^2_F\alpha\omega^6/m^2_e.
\;\;
\label{6}
\end{equation}
 If we compare the two lepton loop to the one lepton loop result we find
\begin {equation}
 T^{(2)}_{pv} / T^{(1)}_{pv} \sim(10^{-4}) G_F M^2_W[ M^2_W/m^2_e].
\;\;
\label{7}
\end{equation}

This ratio is quite large and independent of the photon energy below
$m_e\sim 500keV$; its value is approximately $2\cdot 10^5$. Physically this reflects
the fact that the two final neutrinos are separated in the two loop
Feynman graph by an intermediate electron and a neutrino, and hence a
distance of an electron Compton wavelength. Even for photon energies
above $1 MeV$ the corresponding graph with a muon loop enhances the
amplitude for photon- muon-neutrino scattering by a factor of  $5$.  Should
such an effect be observed it would indicate a violation of $CPT$ and
Lorentz invariance  \cite{9}. Attempts to observe such an effect have been
unsuccessful  \cite{10}.
 Why is the situation different with the parity violation in the forward
amplitude in the scattering channel? The answer is simple. This
amplitude must be odd under parity and therefore be at least cubic in $kp$. So we need an extra factor of $kp$ and this comes (in the one-loop
case) with its own denominator $M^2_W$, for dimensional reasons. Therefore
the parity violating amplitude even in the one loop diagram is formally
of order $G^2_F$ and therefore on par with the two loop diagram. However it
should be emphasized that even with this large enhancement factor the
effect remains extremely small.
 
For real photons $k^2 = 0$,  the two loop estimates are as explained
above, dominant. Therefore estimates of the rotatory power of the
neutrino sea in the literature   \cite{3} have to be multiplied by a factor of
$2\cdot 10^5$ to make them right. In particular, using the Equation (31) of
Abbasabadi and Repko  \cite{3}, and applying this correction we get for the
rotatory power  $\phi /L$ the formula:

\begin {equation}
 \phi/L = 0.015 G^2_F\alpha \omega^2T^2_{\nu}(N_{\nu}  - N_{\bar\nu})/m^2_e
 \;\;,
\label{8}
\end{equation}
which should only apply in the range discussed by Abbasabadi and Repko  \cite{3} ( where $T_{\nu}$ is the neutrino Fermi energy).

For off-mass-shell photons  the situation is different, since the
Landau-Yang theorem does not apply to off-mass shell photons. Therefore even the lowest triangle graph is viable as was realized by Nieves et al \cite{3}.  At low photon energy their one loop estimate remains dominant while at higher photon energies the two loop estimate dominates. The transition between
Low and High photon energies $\omega^{*}$, is given by the equation $\omega^{*} E_{\nu}  > 30 eV^2$, where $E_{\nu}$ is the neutrino energy in the rest frame of the sea. For a
standard sea of energy of a few degrees Kelvin the transition occurs in
the $\omega \sim 100 KeV $region, but for an unconventional sea of Fermi Energy $100 eV$ \cite  {11}
the two loop estimate will dominate even for visible photons. For radio waves the one loop estimate dominates for virtual photons.

\section{Effective Lagrangian}

For a systematic study of neutrino-photon interaction it is useful to work with effective Lagrangians.

The simplest example of the Effective Lagrangian is the four-fermion interactions of neutrinos $\nu$  with electrons $e$.
For fermions with momenta much smaller than intermediate bosons mass one can integrate degrees of freedom associated with $W$ and $Z$  and write the Effective Lagrangian only for fermionic degrees of freedom:

\begin{equation}
 L_{eff} = \frac{G_F}{\sqrt{2}} ( \bar{\nu} \gamma_{\alpha}\nu)( \bar{e}
 \Gamma_{\alpha} e)
\;\; , \label{1}
\end{equation}
where $\Gamma_{\alpha} = g_V \gamma_{\alpha} +  g_A \gamma_{\alpha}\gamma_5$.
In the Standard Model ($SM$) one finds  $g_V = \frac{3}{2} - 2 \sin^2{\theta_W}, g_A = \frac{3}{2}$.
For small momenta this effective Lagrangian is as good as the fundamental Lagrangian of the $SM$.

Consider now the process $\nu(p) + \gamma(k) \to \nu(p) + \gamma(k)$ at momenta smaller than intermediate bosons mass
or even smaller than the mass of electron $m_e$.
Interaction between neutrino and photons occurs only 
through the interaction of photons with charged virtual particles in  loops. Loop diagrams are numerous and a bit complicated. For small momenta $( {pk}/{m^2_W}) \ll {1}$ one can expand $\nu(p)
+ \gamma(k) \to \nu(p) + \gamma(k)$ amplitudes in the power series
in this small parameter. The lowest terms of this expansion can be
represented as a matrix element of the appropriate operators i.e.
of the particular terms of the Effective Lagrangian. 

Each term of the Effective Lagrangian  has to be Lorenz-invariant combination of gauge-invariant electromagnetic field tensor $F_{\mu\nu}$ and left-handed neutrino field $\nu_L =\frac{1}{2} (1 + \gamma_5)\nu$ and their derivatives. Effective Lagrangian has dimension four: $[L]\sim{[m]^4}$. The operators may have higher dimensions $D$.
To preserve correct dimension the coefficients in front of these operators should be proportional
to appropriate power of $(1/m)$, where $m$ is the scale of mass walking inside the loops. 
The actual calculation of the diagrams
gives numerical coefficient in front of the operator.

This line of reasoning is very similar to the naive dimensional arguments given in the Introduction. The only advantage of effective Lagrangian is that within that
more advanced approach  we get more clear understanding of the structure of the operators (i.e. of the scattering amplitudes).  

\subsection{P-even scattering amplitude}
\subsubsection {One-loop approximation}
  Consider effective Lagrangian for $P$-even $\nu\gamma$-scattering. The photon's part of amplitude should be even under parity, i.e. be the same for right-handed photons and left-handed photons. One has to construct
appropriate Lorenz invariant operators from the fields
$F_{\mu\nu}$ and $\nu_L$. The combination of the fields that satisfies all these conditions looks as follows:

\begin{equation}
 L_{eff} \sim  \frac{e^4}{m^4}[  F_{\mu\alpha}F_{\mu\beta}]  \bar{\nu} \gamma_{\alpha}\partial_{\beta}( 1 + \gamma_5) \nu + h.c.
\;\; , \label{10}
\end{equation}

It has dimension $D=8$.
Matrix element of $L_{eff}$  for forward scattering gives the amplitude

\begin{equation}
 T \sim  \frac{e^4}{m^4}( pk)^2 \epsilon(k)^{*}\epsilon(k)
\;\; , \label{11}
\end{equation}

If we identify parameter $m$ in eq.(11) with the largest mass in diagrams (i.e. with $m_W$) we reproduce well known result   \cite{7} up to the numerical constant.

From this exercise it is absolutely clear that non-zero 
$\gamma\nu$ scattering appears only in the second order in photon momenta, i.e. in order $(pk)^2$. Thus one immediately concludes that any results of zero order in $k$, i.e. of order $(G_F \alpha)\sim\alpha^2/m_W^2$ (e.g. such as in  ref.  \cite{ 1, 2}), are
erroneous. 

There is no way to violate Gell-Mann theorem within
effective Lagrangians. To get the amplitude of the order
$G_F\alpha$ one needs operator with $D=6$. By direct inspection one finds that there is no gauge invariant operator with $D=6$.

\subsubsection{Two-loop approximation}

Consider now two-loop amplitudes with light particles (i.e. electrons and neutrino) in intermediate state between two external neutrino vertices. We expect that these diagrams are proportional to $G^2_F \alpha/m^2_e$ . Thus to preserve correct dimension in effective Lagrangian we need operator of dimension $D=10$. Appropriate effective Lagrangian is    

\begin{equation}
 L_{eff} \sim \frac{G^2_F \alpha}{m^2_e}  [ F_{\mu\alpha}(\partial_{\gamma} F_{\mu\beta})]
[ \bar{\nu} \gamma_{\alpha}\partial_{\beta}\partial_{\gamma}( 1 +
\gamma_5) \nu] + h.c. \;\;  \label{12}
\end{equation}

For this $L_{eff}$ the scattering amplitude is of the third order in $(p k)$

\begin{equation}
T \sim C \frac{G^2_F \alpha}{m^2_e} (p k)^3 \epsilon(k) \epsilon^*(k).
\;\;
\label{13}
\end{equation}

 Thus for P-even scattering second order loops give  correction of the order $(G_F M^2_W)(pk/m^2_e)$, i.e. small correction to the one-loop result. 

\subsection{Optical activity. P-odd scattering amplitude}

Now let us come back to $P$-odd effects in $\nu\gamma$ scattering
and find appropriate operators in $ L_{eff}$  responsible for
optical activity. The Lagrangian of dimension $D=8$ that depends on $P$- odd
combinations of photon polarizations has the form

\begin{equation}
 L_{eff} \sim \frac{1}{m_W^4} [ F_{\mu\alpha}\tilde F_{\mu\beta}]
 [ \bar{\nu} \gamma_{\alpha}\partial_{\beta}( 1 + \gamma_5) \nu] + h.c.
\;\; , \label{14}
\end{equation}
where $\tilde F_{\mu\nu} = \frac{1}{2}\epsilon_{\mu\nu\alpha\beta} F_{\alpha\beta}$.

The surprise is that this operator does not work in our case. On
can check that the matrix element of $F_{\mu\alpha}\tilde
F_{\mu\beta}$  between photons with the same momenta and
polarization ( forward scattering) is identically zero and this operator of $D = 8$ does not contribute into $P$-odd
forward scattering. Thus P-odd effects are zero in $(pk)^2/m_W^4$ order. The first nonzero effect is of the third order  $\sim (pk)^3$. Birefringence of neutrino sea is strongly suppressed!

 To find effect in the next order we have to look for the operators of higher dimension $D=10$.
One of these operators looks like follows

\begin{equation}
 L_{eff} \sim \frac{1}{m^6}  [ F_{\mu\alpha}(\partial_{\gamma}\tilde F_{\mu\beta})]
[ \bar{\nu} \gamma_{\alpha}\partial_{\beta}\partial_{\gamma}( 1 +
\gamma_5) \nu] + h.c. \;\;  \label{15}
\end{equation}

With this $L_{eff}$ the forward scattering amplitude for a
photon with momentum $k$ and for a neutrino with momentum $p$ is equal to

\begin{equation}
T = C(e^4/8{\pi}^2 )(pk/m^2)^2\epsilon_{\mu\nu\alpha\beta}
\epsilon_{\mu}(k) \epsilon_{\nu}^*(k) (p_{\alpha} k_{\beta}/m_W^2).
\;\;
\label{16}
\end{equation}
This amplitude has different contribution to left-handed and to right-handed photons scattering : $T_{LL}= -T_{RR}$.
\section{Actual calculations}

\subsection{One-loop calculations. Real photons}

The actual calculation of the coefficient C has been done in one
loop-approximation in \cite{3} with the results
\begin{equation}
T = C(e^4/8{\pi}^2 s^2)(pk/m_W^2)^2\epsilon_{\mu\nu\alpha\beta}
\epsilon_{\mu}^(k) \epsilon_{\nu}^*(k) (p_{\alpha} k_{\beta}/m_W^2)
\;\; ,
\label{17}
\end{equation}
where
\begin{equation}
C  = 4/3 ( \ln(m_W ^2/m^2)-11/3 ),
\;\;
\end{equation}
 in the third reference in \cite{3} and
\begin{equation}
C  = 4/3 ( \ln(m_W ^2/m^2)-8/3 ),
\;\;
\end{equation}
in the fourth one  \cite{3}. The reason for that discrepancy is unknown. Though numerically eqs.(18)  
and (19)  differ by only a few per cent it would be interesting to understand whether there is a correct one-loop calculation.

On the other hand we find that for a $P$-odd effect two-loop diagrams by many order of magnitude larger than one-loop contribution. Thus we can neglect any one-loop results.

\subsection{One-loop calculations. Off-shell photons}

 The optical activity for off-shell photon was first considered by Mohanty, Nieves and Pal in \cite{3}. They  noticed that Gell-Mann prohibition theorem does not work for the off-shell photons. Thus one can expect that off-shell amplitude is of the first order in $1/m^2_W$. Indeed

\begin{equation}
 T = (e^4/8{\pi}^2 s^2)\epsilon_{\mu\nu\alpha\beta}
\epsilon_{\mu}^(k) \epsilon_{\nu}^*(k) (p_{\alpha} k_{\beta}/m_W^2)
(k^2/6m_e^2) \;\;,
\label{20}
\end{equation}
where $s=sin \theta_W$.
 
Eq. (20) differs from the original result of ref. \cite {3} by factor $1/2$. The reason is that triangle diagram was missing there. (This is the triangle diagram with $Z$ boson and two photons. For real photons, $k^2=0$ the
triangle diagrams cancel each other in the SM since the SM is anomaly free. But for off-shell photons each triangle
diagram gives a contribution proportional to $k^2/m^2$. The sum of all triangles is nonzero and the main contribution comes from the electron loop. This contribution has to be taken into account)

\subsection{Two-loop calculation}

The physical reason for the dominance of the two-loop diagrams under the one-loop in P-odd amplitudes is simple.
To escape  Gell-Mann's restriction one needs non-local
interactions in order to include higher orbital momenta of the
pair $\nu\bar{\nu}$ into annihilation process. In the one loop
approximation the expectation value of the orbital momentum of the neutrino pair is $\sim {p}/{m_W}$.The factor ${1}/{m_W}$  measures the shortest separation of two neutrino during interaction (non-locality).

In the two-loop approximation the expectation value of neutrino pair orbital moment is
$\sim {p}/{m_e}$. The factor ${1}/{m_e}$ is due to  $(e^-e^+\nu)$  in the intermediate states.

 Kinematically P-odd amplitudes are of the order of $(pk)^3$ (see section 2.2). On dimensional ground we conclude that  two-loop amplitudes are of the same order  $(pk)^3$ (see secton 2.1.2). Thus one-loop P-odd amplitude has the same dependence on 
$(pk)$ as the two-loop P-odd amplitude.
Moving to the next order in electro-weak interaction
we loose a small factor $\alpha/2\pi$ , but win a great factor
$(m^2_W/m^2_e)$. The net effect is

\begin{equation}
(T^{(2)}/T^{(1)}) \sim (\alpha/2\pi)(m^2_W/m^2_e) \sim 10^{7}.
\;\;
\label{21}
\end{equation}

Actual calculation is rather lengthy. The result is

\begin{equation}
T^{(2)} = (13/27)(G^2e^2/64{\pi}^4)(g^2_V + g^2_A)(pk)^2\epsilon_{\mu\nu\alpha\beta}
\epsilon_{\mu}(k) \epsilon_{\nu}^*(k) (p_{\alpha} k_{\beta}/m_e^2).
\;\;
\label{22}
\end{equation}

Thus the enhancement factor is

\begin{equation}
T^{(2)}/T^{(1)} = (\alpha/64\pi sin^2\theta_W)(m^2_W/m^2_e)(13/27C)(g^2_V + g^2_A)\sim  10^{5},
\;\;
\label{23}
\end{equation}
where $C$ is one-loop coefficient from eq.(8). We have lost two
order of magnitude compared with naive estimate in eq.(10)  mainly due to the
large logarithmic coefficient $C$ in one-loop amplitude.
Still enhancement factor is very large $\sim 10^5$ !

This research was supported by NSERC-Canada and by RFBR
grants 00-15-96562.



\begin{thebibliography}{99}



\bibitem{1} H.Y.Chiu and P.Morrison, Phys. Rev. Lett. {\bf 5} (1960) 573;
S.G.Matinyan and N.N. Tsilosuni, Sov.Phys. JEPT {\bf 14}
(1961) 1195.  
 
\bibitem{2} J.Royer, Phys. Rev. {\bf 174} (1968) 1719;
L.Stodolsky, quoted in  G.Karl, Canadian Journal of Physics {\bf 54} (1976) 568.

\bibitem{3} see e.g.: S.Mohanty, J.F.Nieves and P.Pal, Phys. Rev. {\bf D58} (1998) 093007; Ali Abbasabadi and Wayne W.Repko, Phys.Rev. {\bf D64}(2001)113007; ibid, {\bf D67} (2003) 073018; G.Karl and V.A.Novikov, hep-ph/0009012; V.B.Bezerra et al, Phys.Rev {\bf D67} (2003) 084011; V.B.Bezerra, C.N.Fereira,J.A.Helayel-Neto, hep-th /0405181.
 
\bibitem{4}  R.E.Peierls: "Surprises in Theoretical Physics" Princeton University Press,Princeton, NJ 1975

\bibitem{5} M.Gell-Mann, Phys. Rev. Lett. {\bf 6} (1961) 70.

\bibitem{6} L.D.Landau, Dokl. Akad. Nauk USSR {\bf60}, 207 (1948); C.N.Yang, Phys. Rev. {\bf77}, 242 (1950)

\bibitem{7} M.J.Levine, Nuovo Cimento {\bf A48} (1967) 67.
 
\bibitem{8} Duane A. Dicus and Wayne W.Repko, Phys.Rev. {\bf D48}(1993) 5106

\bibitem{9} S.Coleman and S.Glashow, hep-ph/9812418;
 R.Jackiw and V.A.Kostelecky, Phys. Rev. Lett. {\bf 82} (1999) 3572
 
\bibitem{10} see e.g. J.N.Clarke, G.Karl and P.J.S.Watson, Canadian Journal of Physics {\bf 60}
(1982) 1561.
 
\bibitem{11} V.M.Lobashev et al, Phys. Lett. {\bf B460} (1999) 227.
\end{thebibliography}
\end{document}